# Polarization and polarization induced electric field in nitrides – critical evaluation based on DFT studies

*Pawel Strak*[*], *Pawel Kempisty*[*], *Konrad Sakowski*[*], *and Stanislaw Krukowski*[*,†]

[*]Institute of High Pressure Physics, Polish Academy of Sciences, Sokolowska 29/37, 01-142 Warsaw, Poland

[†]Interdisciplinary Centre for Materials Modeling, Warsaw University, Pawinskiego 5a, 02-106 Warsaw, Poland

**ABSTRACT**

It was shown that polarization could be defined only for finite slabs with specified boundary condition. In the case of nitrides in wurtzite structure two different polarizations occurred which differ by magnitude and direction of built-in electric field. Density Functional Theory (DFT) calculations were used to evaluate polarity of group III nitrides, such as aluminum nitride (AlN), gallium nitride (GaN) and indium nitride (InN). Two different approaches to polarization of nitride semiconductors were assessed. It was shown that Berry phase formulation of the electron related polarization component provides nonzero polarization even for single atom and additionally a number of various solutions, different for various selection of the simulated volume of the nitrides. The electronic part gives saw-like pattern for polarization. Alternative standard dipole density formulation of polarization depends on the selection of the simulation volume in periodic continuous way. A condition of continuous embedding into the infinite medium, and simultaneously, the zero surface charge representation at crystal boundary provides to physically sound solution. These solutions correspond to maximal and minimal polarization values. These solutions arise from different physical termination of the crystal surfaces, either bare or covered by complementary atoms. This change leads to polarization and electric field reversal. The polarization and related built-in electric fields were obtained.

## I. INTRODUCTION

Macroscopic polarization of crystalline materials is physically important property of solids[1]. Polarization is related to symmetry of a crystalline lattice, which allows for the existence of vector property, in the point group symmetry of the lattice. It is therefore expected that microscopic definition of the property is formulated, in direct relation to the atomic lattice structure. Unfortunately typical statement, frequently encountered in textbooks, defining polarization as dipole of a unit cell, divided by its volume, was criticized[2-4]. It was claimed that so defined quantity depends on the selection of the simulation volume so it could not be used to determine the physical property of the matter. The statement was related to the presence of the surface term, inherently involved in the definition of the polarization as an electric dipole density, which affects its value[5-9]. A direct modification, such as using of large volume, does not remove this deficiency[2]. It is still not clear whether additional requirement of independency of the selection of simulation volume should be applied to determination of the polarization. Recent results shed serious doubts whether such requirement could be met for infinite systems[10-12].

Macroscopic polarization is technically important physical property of pyro- and ferro-electrics, often used in many important technological applications. A natural use of permanent or induced electric dipole could be found in many applications[13]. Natural consequence of polarization is presence of an electric field in the crystal interior possibly affecting functionality of advanced electronic and optoelectronic devices. The built-in electric fields affect energy of quantum states that is known as Stark effect. In addition, strain induced field may contribute to this effect significantly, especially in strained quantum low dimensional structures frequently used in modern devices, built on polar GaN(0001) surface[14-17]. The electric field changes energy of quantum states of both types of carriers, electrons and holes, giving rise to phenomenon for long time known as Quantum Confined Stark Effect (QCSE) [17]. A mere change of the energy of quantum states could be either beneficial or harmful; a really detrimental is spatial separation of electrons and holes that are shuffled to the opposite ends of the quantum well[17-19]. Spatial separation reduces an overlap of the hole-electron wavefunctions, their radiative recombination rates and lowers efficiency of photonic devices[20-23]. The negative influence of QCSE may be enhanced by Auger recombination or carrier leakage at high injection currents[24-28]. That could lead to decrease of the device efficiency for higher injection currents, the phenomenon nicknamed as "efficiency droop"[29]. A harmful influence of QCSE for optoelectronic devices is compensated by its beneficial contribution on electronic devices based on two-dimensional electron gas (2DEG), such as field effect transistors (FET's) or molecular sensors. Electric field at AlN/GaN heterostructures stabilizes 2DEG leading to high carrier mobility which may be used for construction of fast electronic devices. The electric field, induced by dipoles of the molecules, attached to the surface, may contribute to the sensitivity of molecular sensors which opens new applications of such devices.

Polarization is therefore a physical property that is an increasingly important in technology. Recently formulated approach was to divide polarization into ionic and delocalized charge and to calculate the change of polarization only[8, 9]. The ionic part may be calculated directly, the delocalized contribution may be obtained using Berry phase formulation[2-7]. This procedure provides required quantity modulo some factor. In the work presented below we will critically asses this formulation and compare these results with reformulated standard definition of polarization as dipole density with additional condition imposed at boundaries. These conditions allow incorporation of the vacuum in the infinite crystal body and let to avoid generation of surface charge. Thus polarization is determined exactly without any inference from the boundary terms. As it is also presented below, polarization induced electric field in the bulk solid may be determined. That determination is compatible with the selection of the termination surfaces, thus provide base for accounting of the field influence on the properties of optoelectronic and electronic devices.

## II. CALCULATION METHODS

In the calculations reported below three different DFT codes were used: commercially available VASP[30-32], freely accessible SIESTA[33-35] and commercially accessible Dmol[36]. In the first instance a standard plane wave functional basis set, as implemented in VASP with the energy cutoff of 29.40 Ry (400.0 eV), was used. As was shown by Lepkowski and Majewski, it gives good results in precise simulations of GaN properties by VASP code[37]. The Monkhorst-Pack grid: (7x7x7), was used for k-space integration[38]. For Ga, Al, In and N atoms, the Projector-Augemented Wave (PAW) potentials for Perdew, Burke and Ernzerhof (PBE) exchange-correlation functional, were used in Generalized Gradient Approximation (GGA) calculations[39-41]. Gallium 3d and Indium 4d electrons were accounted in the valence band explicitly. The energy error for the termination of electronic self-consistent (SCF) loop was set equal to $10^{-6}$. The obtained lattice constants were: GaN - a = 3.195 Å and c = 5.206 Å, AlN - a = 3.112 Å and c = 4.983 Å, and InN - a = 3.563 Å and c = 5.756 Å, which is in good agreement with the experimental data: GaN: a = 3.189 Å and c = 5.185 Å, for AlN: a = 3.111 Å and c = 4.981 Å and for InN: a = 3.537 Å and c = 5.706 Å.

The second code, SIESTA uses norm conserving pseudopotentials with the numeric atomic orbitals local basis functions that have finite size support, determined by the user[33-35]. The pseudopotentials for Ga, Al, In and N atoms were generated, using ATOM program for all-electron calculations[42,-43]. Gallium 3d and Indium 4d electrons were included in the valence electron set explicitly and were represented by single zeta basis. For s and p type orbitals quadruple zeta basis set were used. Aluminum atom basis was represented by triple zeta function. Integrals in k-space were performed using 3x3x3 Monkhorst-Pack grid. The minimal equivalent of plane wave cutoff for grid was set to 275 Ry. As a convergence criterion terminating SCF loop, the maximum difference between the output and the input of each element of the density matrix was employed being equal or smaller than $10^{-4}$.

For comparison, spontaneous polarization was also calculated using DMol3 commercial program[36]. In DMol3 package, full-electron Kohn-Sham eigenvalue problem with periodic boundary conditions (PBC) for wave function is solved, using basis of Linear Combination of Atomic Orbitals (LCAO) [44]. The double zeta plus polarization (DZP) basis set was applied, in which required matrix elements are evaluated numerically on the properly chosen grid. For exchange energy, approximation proposed by Becke[45] and Lee, Yang, Parr[46] (B88) was chosen, for correlation energy Tsuneda, Suzumura, Hirao functional was used[47]. The electric potential was determined by solution of Poisson equation which was expressed as sum of the multipole contributions, centered on each atom of the simulated volume with the optimized cutoff radii[36, 48]. In order to account long range interactions Ewald summation is employed[49]. The method is verified to assure solution tolerance of $10^{-6}$ hartree. Thus the DMol3 Poisson equation solution procedure is different than that of VASP and SIESTA.

## III. DENSITY DISTRIBUTION

The electric fields in VASP and SIESTA were obtained by inverse Fast Fourier Transform (FFT) method based on periodicity of electrostatic potential at the edges of the simulated volume. Such periodicity is enforced by implicit addition of appropriate external field so that any potential changes related to polarization are compensated. DMol3 solution is based on multipole expansion method, employing Ewald summation procedure to account long range interaction. Since the DMol3 wavefunctions basis, are spherical harmonics, that are perfect electric multipoles, the summation employs coefficients only, assuring extremely fast convergence of the Poisson equation solution procedure. Siesta uses both FFT method and molecular orbitals basis set while VASP is combination of FFT and planewaves, therefore such combination of the codes allows exhaustive verification of these three approaches. In Fig. 1 the c-plane averaged density profiles were presented for AlN, GaN and InN. The diagrams presents the electronic density arising from simple superposition of atomic charges $\rho_{SAO}$, the density obtained by full solution of DFT Kohn-Sham equation $\rho_{KS}$ and the difference of these two:

$$\Delta\rho \equiv \rho_{KS} - \rho_{SAO} \qquad (1)$$

Naturally, any superposition of charge of separated atoms has no electric dipole moment, therefore either Kohn-Sham density $\rho_{KS}$ or the density difference $\Delta\rho$ may be used for calculation of the polarization and the polarization induced electric field. In principle the results obtained using both approaches should differ only by an error arising from intersection of the subtracted separated atom charge (i.e. effectively positive charge, representing atomic nuclei) by top and bottom boundaries. The coordinate shift, resulting from the periodic boundary conditions may affect magnitude of the dipole obtained from the integration over simulated volume.

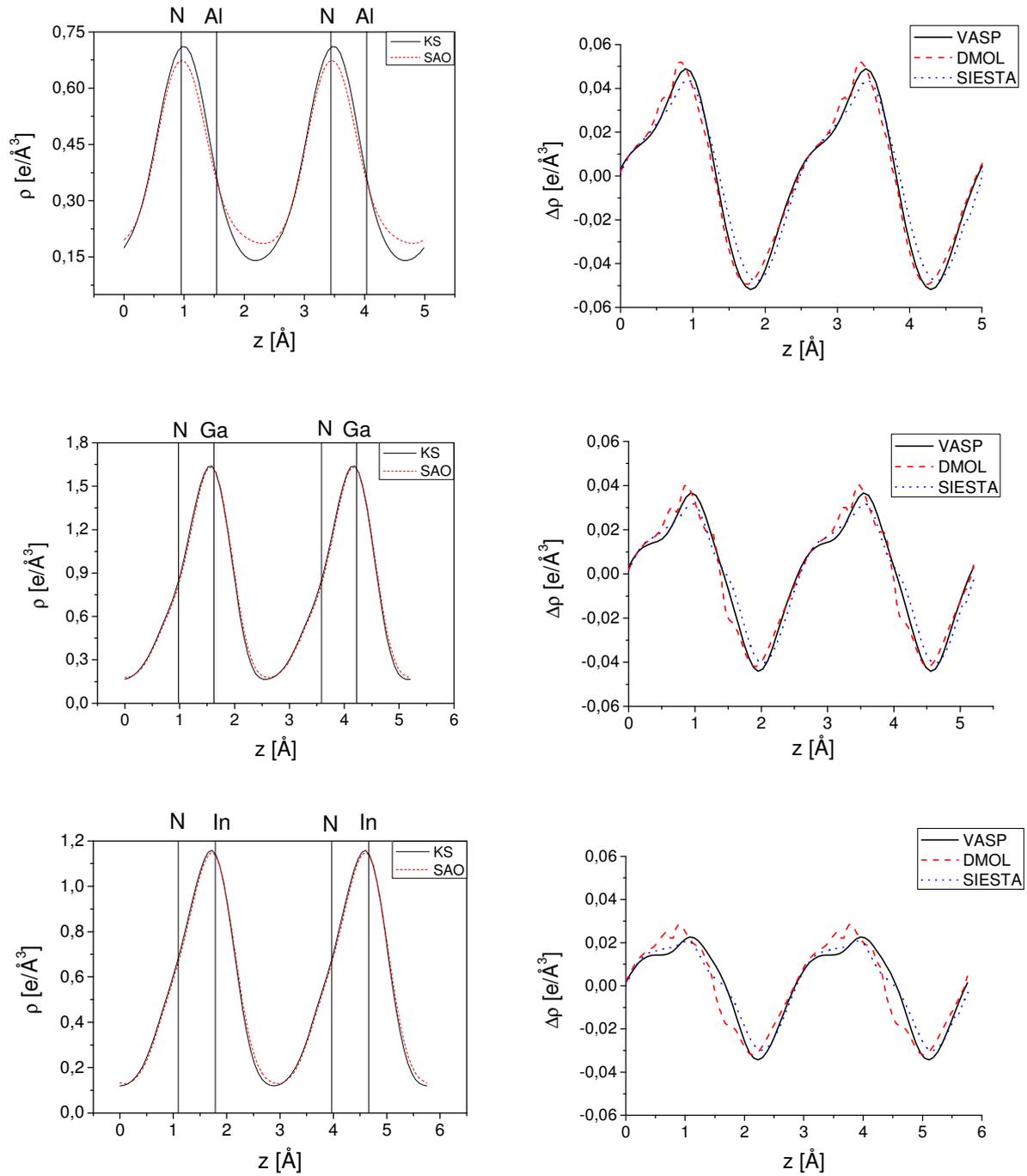

Fig. 1. (Color online) Left column presents superposition of individual atom charges density ($\rho_{SAO}$ - red dashed line) and DFT Kohn-Sham solution ($\rho_{KS}$ - black solid line) obtained by VAPS code. Right column presents the density difference $\Delta\rho$. The densities are averaged in the plane perpendicular to c-axis diagrams, along which they are plotted for: AlN(top); GaN(middle); InN(bottom).

It is worth noting that the density difference is relatively small. It is remarkable that these solutions are close for all three methods. That indicates on good convergence of the calculations and proper representation of the real density distribution by DFT solution.

Boundary conditions enforce periodic density distribution which does not specify simulated volume entirely as its boundaries can be selected arbitrarily. Since the simulated system is electrically neutral, integral over the density should vanish, thus there are at least two different locations of the boundaries that of the zero averaged density difference. In principle any other choice could be adopted, giving rise to different values of the dipole moment of the sample, and consecutively different value of polarization. As it is shown below, the appropriate treatment of the boundary problem allows us to obtain well defined, physically sound magnitude of the dipole and consequently, the polarization.

## IV. POLARIZATION OF BULK AlN, GaN AND InN

Polarization in the solids and in the molecules arises from the electronic charge transfer resulting from bonding that leads to emergence of electrical dipoles. The polarization may be obtained from its definition, being equal to the dipole density:

$$\vec{P} = \frac{1}{\Omega} \int d^3 r \, \rho(r) \, \vec{r} \tag{2a}$$

where $\Omega$ is the simulated volume, $\rho(r)$ charge density, both electronic and nuclei. Naturally polarization of superposition of charges of individual atoms should have polarization equals zero, i.e.

$$\vec{P} = \frac{1}{\Omega} \int d^3 r \, \rho_{SAO}(r) \, \vec{r} = 0 \tag{2b}$$

For the relation (2b) fulfilled, the polarization may be obtained equally from Kohn-Sham density $\rho_{KS}$ or the density difference $\Delta \rho = \rho_{KS} - \rho_{SAO}$, which merely reflects above stated fact that it arises from electronic charge transfer:

$$\vec{P} = \frac{1}{\Omega} \int d^3 r \, \rho_{KS}(r) \, \vec{r} = \frac{1}{\Omega} \int d^3 r \, \Delta\rho(r) \, \vec{r} \tag{2c}$$

It was recognized that for electrically neutral systems the result of such procedure does not depend on the choice of the coordinate system, nevertheless it depends on the choice of the simulated area[4, 6, 7]. It was a subject of long debate whether polarization may be uniquely defined as bulk properties of the solids, independent of their surfaces[8,9]. A simple model presented in Fig. 2. show that the polarization depends on the boundary conditions, even in the simplest case for the charge transfer represented by point charges.

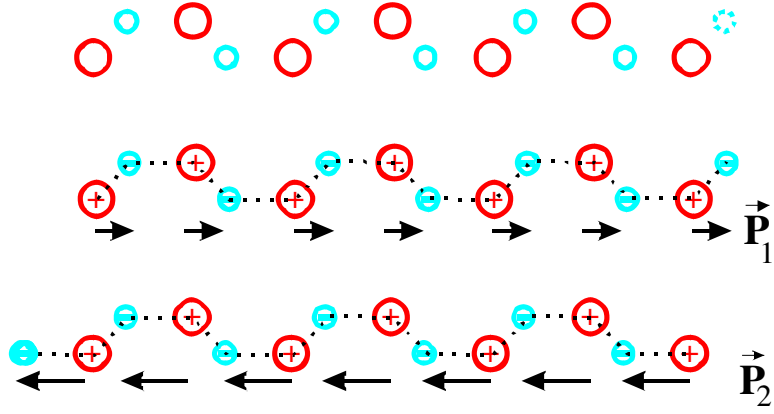

Fig. 2. (Color online) Chain of metal (Al, Ga, In – large red circles) and N (small blue circles) atoms, representing polarization properties of nitride lattice. Top – superposition of individual atoms, electrically neutral, for which both polarization and average electric field vanish, center and bottom – two polarized (charged) lattices, differ by shift of the edge metal atom (bonding is denoted by dotted line). The edge atom, i.e. the one shifted, is represented in the top diagram by broken line. Polarization is represented by black arrows.

As it is shown in Fig. 2, at least two equivalent polarization values could be defined, depending on the termination of lattice, i.e. on the position of the edge atoms. These two polarizations differ by both magnitude and the direction. Therefore, the definition of the polarization as purely bulk property, without reference to termination of the lattice cannot by physically justified. Polarization of the infinite medium cannot be uniquely determined, as at least the two different polarization values exist for the infinitely thick slab. The difference is not related to calculation method, it is physical in nature, as different electric field in the medium arises due to different boundaries. Therefore the polarization could be defined in finite slab only. In addition, the procedure determining polarization and the polarization related field has to enforce zero surface charge and zero external field for these two quantities, respectively.

More recent approach is based on division of the polarization into the ionic and electronic contributions:

$$\vec{P} = \vec{P}_{ion} + \vec{P}_{el} \qquad (2)$$

where ionic contribution is treated in standard manner using summation over all point charges of the nuclei $Ze$:

$$\vec{P}_{ion} = \frac{1}{\Omega} \sum Ze \cdot \vec{r} \qquad (3)$$

and Berry phase formulation is applied for electronic part, based on linear response theory adiabatic change of the potential, controlled by parameter $\lambda$:

$$\Delta \vec{P}_{el} = -\frac{2e}{\Omega} \sum_n \int \vec{r} \left( \left| w_n^1 \right| - \left| w_n^0 \right| \right) d^3 r \qquad (4)$$

where the sum runs over all occupied real space Wannier functions $\left| w_n^0 \right|$ and $\left| w_n^1 \right|$, calculated for both terminations of the adiabatic path (superscripts correspond to $\lambda = 0$ and $\lambda = 1$, respectively) [4-7, 49].

## A. Polarization obtained from Berry phase formulation

Berry phase formulae for polarization, given in Refs 4-7, modified for the application to USPP's and PAW datasets[50], were used in the version implemented in VASP package. In order to obtain good approximation for band gap we have recalculated wavefunctions for PBE charge density with HSE03 functional[51]. Determination of polarization from geometric phase formulation relies on adiabatic change of the crystal potential. Derivation of polarization by Resta is limited to electronic part only[2]. The ionic and electronic contributions, obtained from Eq. 4 and 5 respectively without assumption of electric neutrality, depend on the coordinate system and the truncation of the integration area. As an initial test, the Berry phase polarization was determined using single Ga atom in the 20 Å long cell.

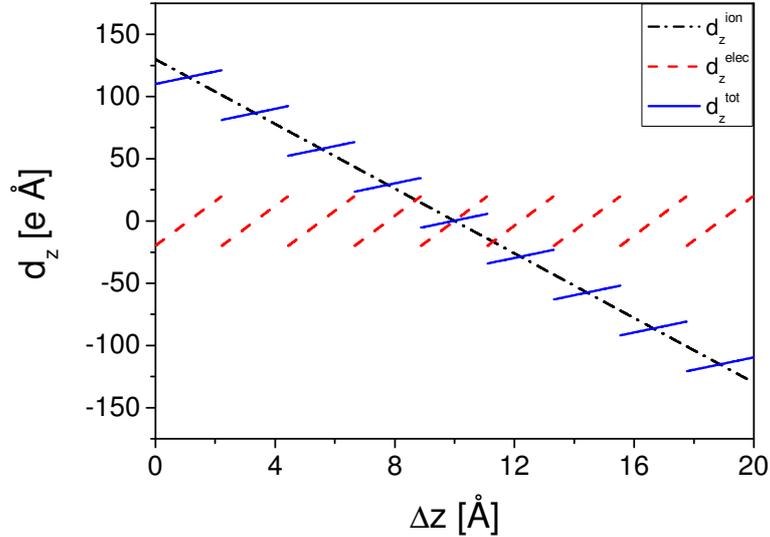

Fig. 3. (Color online) Electric dipole of single gallium atom located in 20 Å long simulation cell as a function of the shift of the periodic cell related to the Ga atom along z-axis obtained from Berry phase formulation implemented in VASP: ionic part (Eq. 4) – black dash-dotted line, electronic part (Eq. 5) – red dashed line, total – solid blue line. $\Delta z = 0$ corresponds to a system with Ga atom located in the center of a cell. $\Delta z = 10$ corresponds to a system with Ga atom located in the boundary of a cell.

As shown in Fig.3. the Berry phase expression is not constant, it depends on the location of single atom in the simulation cell. Next, the solution obtained by the iteration procedure where the density is kept constant, equal to $\rho_{SAO}$. The SCF iteration loop converged to the wavefunction describing nonpolarized state of the system. Naturally, the Berry phase procedure should give the polarization of the system equal to zero. As shown in Fig. 4, the polarization of Ga-N system is not zero. Moreover, there is no such selection of the simulation volume for which the total polarization vanish.

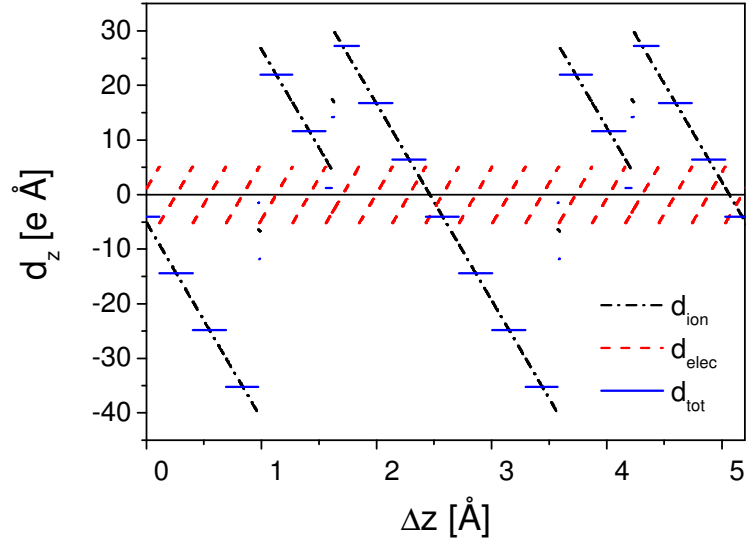

Fig. 4. (Color online) "Z" component of a Ga-N dipole as a function of shift of the periodic cell along c-axis obtained from Berry phase formulation implemented in VASP code: ionic part (Eq. 4) – black dash-dotted line, electronic part (Eq. 5) – red dashed line, total – solid blue line for the wavefunction compatible with the density being the sum of densities of separated atoms $\rho_{SAO}$ of gallium and nitrogen atoms.

Subsequently, the full solution of Kohn-Sham equation was obtained in which the relaxation procedure was performed for all three nitrides: AlN, GaN and InN.

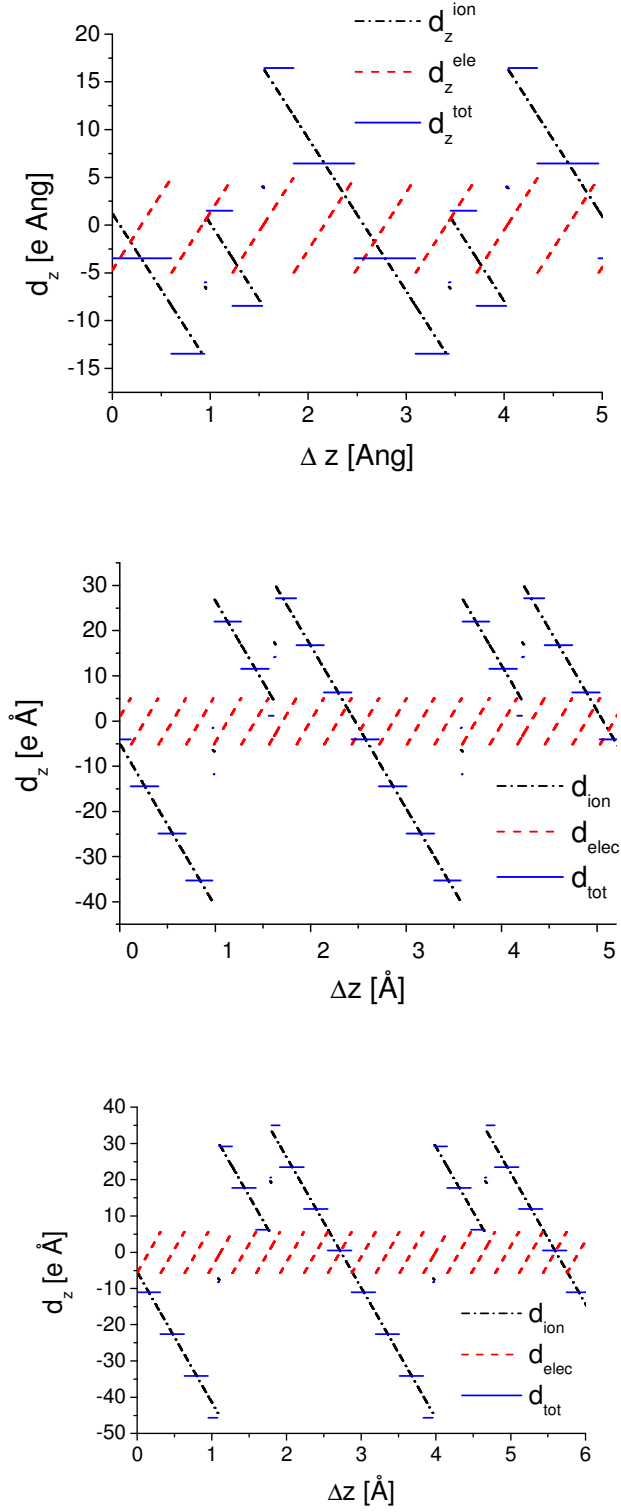

Fig. 5. (Color online) "Z" component of a dipole of simulation cell as a function of shift of periodic cell along c-axis, obtained from Berry phase formulation VASP code: ionic part (Eq. 4) – black dash-dotted line, electronic part (Eq. 5) – red dashed line, total – solid blue line; (top) AlN; (middle) GaN; (bottom) InN.

As claimed by Resta et al, combination of electronic and ionic contributions should be independent of the system coordinates, and also of the truncation of the volume. As expected, the ionic contribution is linear function of the z coordinate that undergoes jumps when the N or Ga atoms cross the boundaries. For 4 atoms in total, four jumps in ionic contribution was obtained for all three nitrides. It was also argued that the electronic contribution is determined modulo periodic change of the geometric phase due to translation of the crystal as a whole, which leads to the following uncertainty:

$$\Delta \vec{P}_{el} = \frac{fe}{\Omega} \sum_n \vec{R}_n \qquad (5)$$

where f is the number of the electrons in valence band and R is the lattice period. Accordingly, the number of electrons is: $f = 8$ for AlN, $f = 18$ for GaN and InN, and the number of the jumps follows this prediction[4]. As expected the total polarization being combination of these two contributions is constant but it contains a number of jumps.

## B. Polarization obtained from dipole density

As it was argued above, the polarization has to be simulated using finite region with appropriate boundary conditions, preventing emergence of surface charge which may affect the results. Nevertheless, in typical ab intio calculations small size cell with periodic boundary conditions for electronic density is used. It is therefore natural to verify whether the periodic boundary conditions affect the determined electric dipole moment. The results of the calculations for single gallium atom related dipole in 20 Å cell are presented in Fig. 6.

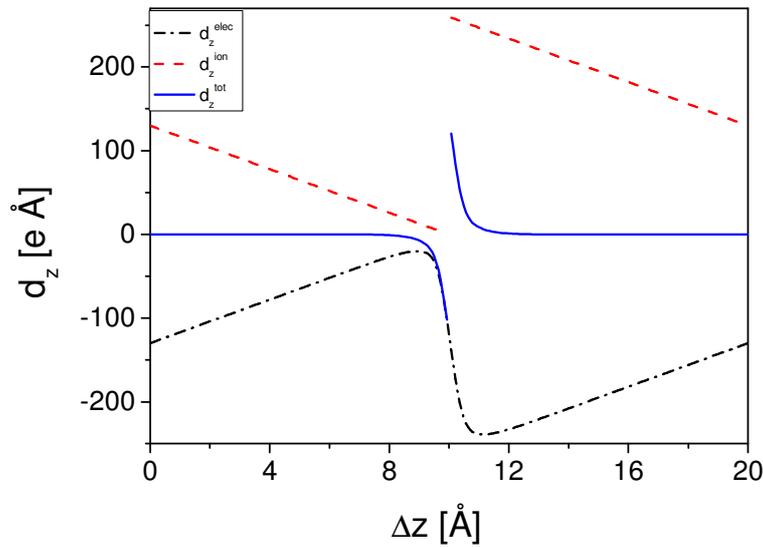

Fig.6 (Color online). Electric dipole of gallium atom in 20 Å long cell, in function of the shift of the periodic cell along z-axis: electronic part – black dash-dotted line, ionic contribution – red dashed line, total – solid blue line. $\Delta z = 0$ corresponds to a system with Ga atom located in the center of a cell. $\Delta z = 10$ corresponds to a system with Ga atom located in the boundary of a cell.

As shown here, the obtained dipole is affected only for these locations where the boundary intersects electronic charge close to Ga atom. Otherwise, the result is zero as expected for nonpolarized system. Similar behavior was obtained for cell simulation of the system of gallium and nitrogen atoms. The dipole related polarization is calculated for the case of density obtained from superposition of atomic charges $\rho_{SAO}$ and solution of Kohn-Sham equation $\rho_{KS}$, presented above in Fig.1.

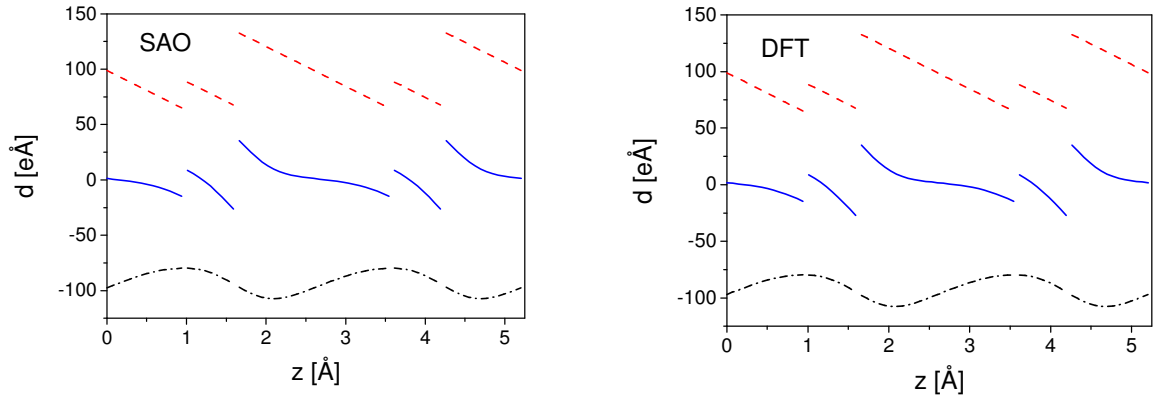

Fig.7. (Color online). Electric dipole of GaN cell in function of the position of the cell boundary: left - determined for density obtained for superposition of atomic charges $\rho_{SAO}$; right - solution of Kohn-Sham equation $\rho_{KS}$: electronic part – black dash-dotted line, ionic contribution – red dashed line, total – solid blue line.

As shown in Fig. 7, direct determination of the dipole is strongly affected by boundary conditions, shifting part of the charge and completely changing the obtained results. In addition, the selection of arbitrary conditions is not compatible with the requirement of the zero surface charge, as the finite density cannot be smoothly terminated, i.e. the consistent method of simulation both the cell representing the volume, and the cell at the boundary of the slab, necessary for simulation of finite systems. For finite slab it is necessary to set zero surface charge in order to prevent additional, spontaneous polarization irrelevant contributions. As there should be no charge or dipole layer at the surfaces, both the electric potential and the electric field are continuous across the system boundaries. In addition, as the modeled sector should be naturally embedded into the infinite medium, the electric potential, the field and its normal derivative should be continuous across the interface, amounting to the following conformity conditions

$$\lim_{\varepsilon \to 0} \phi(z = \varepsilon) = \lim_{\varepsilon \to 0} \phi(z = -\varepsilon) = \phi(0) \quad (6)$$

$$\lim_{\varepsilon \to 0} \vec{E}(z = \varepsilon) = \lim_{\varepsilon \to 0} \vec{E}(z = -\varepsilon) = \vec{E}(0) \quad (7)$$

$$\lim_{\varepsilon \to 0} \frac{\partial E_z(z = \varepsilon)}{\partial z} = \lim_{\varepsilon \to 0} \frac{\partial E_z(z = -\varepsilon)}{\partial z} = \frac{\partial E_z(z = 0)}{\partial z} \quad (8)$$

The considered field could be integrated over the area of the boundary to get average values, which are function of $z$ coordinate only. Since the simulated area can be truncated at any position and the electric field follows the equation:

$$\frac{\partial E_z(z)}{\partial z} = \frac{\rho(z)}{\varepsilon} \quad (9)$$

in which the averaged density profiles, plotted in Fig.1 could be used. Note that the boundary shift should entail appropriate rearrangement of the density distribution.

In order to model the polarization exactly, it is required that the boundaries of the simulated regions should represent the surfaces of real crystal, without additional surface charge. The physically sound approach is that the electronic density difference vanishes outside. Abrupt termination in nonphysical, and any modification of Kohn-Sham density $\rho_{KS}$ to assure continuous could in principle provide additional surface charge. In contrast to that this condition can be imposed for the density difference $\Delta\rho$ so this quantity is used in polarization determination below. Application of the condition (Eq. 9) to assumption of continuity (Eq. 8) entails constant density outside the simulated region as shown in Fig. 6.

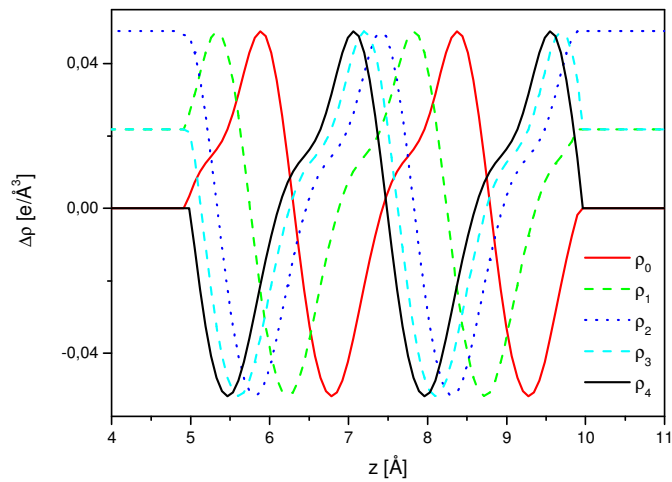

Fig. 8. (Color online) Several choices of the truncation of AlN density difference profiles obtained from VASP and its continuation in accordance to the conformity conditions in Eqs. 6-9.

Therefore, only the two selected terminations, those plotted by solid lines in Fig. 8 for which the density difference is zero outside are suitable for calculation of the polarization. It has to be stressed out that this conclusion results from simultaneous requirement that the embedded region is smoothly incorporated into the crystal body and their edge should represent the surface of real crystal without any surface charge. Due to electric neutrality condition, at least two such terminations could be found for any polarized system.

In fact the difference in the selection of the simulated volume directly affects the resulting value of the electric dipole. Using obtained surface averaged electron density difference profiles, the AlN electric dipole magnitude was calculated directly by integration according to Eq. 2c. The result is presented in Fig. 9.

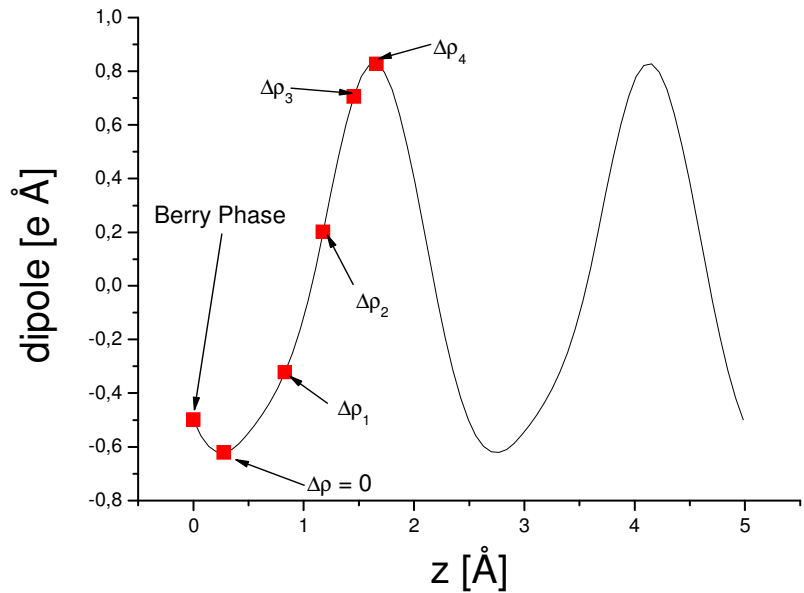

Fig.9. (Color online) Electric dipole of the AlN simulated volume, in function of the shift along c-axis. The results obtained by integration of the profiles truncated in the way presented in Fig. 2 are denoted by red squares.

For real slab zero density condition outside is valid which selects from all possible truncations only the two that are denoted by solid lines in Fig. 8. As expected these two selections correspond to maximal and minimal values of the dipole plotted in Fig. 9. In fact, these two selections correspond to two possible termination of the top polar surface, i.e. by nitrogen or aluminum atoms. These two cases correspond to the top surface having aluminum triply bonded atoms that are either bare or nitrogen covered. In order to fulfill chemical stoichiometry slab criterion, such change of the top surface and position of the layer of N atoms has to be accompanied by the appropriate rearrangement of the bottom surface. In summary this leads to reversal of the polarization dipole, as shown in Fig. 3.

It has to be noted that the modeled area and the interfaces does not describe the real atomic configuration of any nitride surface. This is a simulation model devised to obtain the polarization and electric field in uniformly polarized nitrides $\vec{E}_0$, corresponding to minimal energy of the nitrides under no external field. The magnitude of such field will be determined in the next Section.

### C. Polarization - summary

It was postulated that the Berry phase results should give the physically meaningful value of polarization by proper selection of the additional constant value given by Eq. 10 (Ref. 4). This is not as straightforward as it was supposed to be. The number of the possible selection of arbitrary constant is large, even for simple structure of the nitrides. Naturally, the Berry phase result should be within the interval obtained from dipole calculations. This is not the case, but it could be reconciled by additional contribution which may bring the Berry phase result to this interval. In order to compare these results, the obtained polarization values are summarized in Table I.

Table I. Polarization of nitrides; AlN, GaN and InN, obtained from dipole and Berry phase calculations (in e/Å$^2$)

|             | AlN      | GaN     | InN     |
|-------------|----------|---------|---------|
| Berry phase | -0.3220  | -0.7666 | -0.7205 |
|             | -0.2028  | -0.5404 | -0.5386 |
|             | -0.1432  | -0.3142 | -0.3568 |
|             | -0.0836  | -0.2572 | -0.1749 |
|             | 0.0357   | -0.0880 | -0.1275 |
|             | 0.0952   | -0.0335 | 0.0070  |
|             | 0.1549   | 0.0250  | 0.0980  |
|             | 0.3934   | 0.1382  | 0.1889  |
|             |          | 0.2513  | 0.2798  |
|             |          | 0.3082  | 0.3256  |
|             |          | 0.3644  | 0.3708  |
|             |          | 0.4774  | 0.4617  |
|             |          | 0.5905  | 0.5526  |
| Δρ          | -0.01485 | -0.0128 | -0.0114 |
|             | 0.01979  | 0.0153  | 0.0123  |

These polarization values were obtained using the following cell volumes: $\Omega_{AlN}$ = 41.79 Å, $\Omega_{GaN}$ = 46.03 Å and $\Omega_{InN}$ = 63.30 Å. Using Eq. 10 and the lattice constants $c_{AlN}$ = 4.983 Å, $c_{GaN}$ = 5.206 Å and $c_{InN}$ = 5.756 Å, the following additive constants for polarization were obtained: $\Delta P_{el-AlN}$ = 0.954 e/Å$^2$, $\Delta P_{el-GaN}$ = 2.036 e/Å$^2$ and $\Delta P_{el-InN}$ = 1.637 e/Å$^2$. These values are relatively high compared to the dipole results obtained with method based on density difference Eq. 1. It is worth mentioning that the jumps in the electronic part follow Eq. 10 for f = 1, i.e. accounting single electron contribution only. Such jumps could be translated into the following additional constant values: $\Delta P_{el-AlN-1}$ = 0.1192 e/Å$^2$, $\Delta P_{el-GaN-1}$ = 0.1131 e/Å$^2$ and $\Delta P_{el-InN-1}$ = 0.0909 e/Å$^2$. In fact these polarization values obtained in dipole model may be approximated by appropriate subtraction of single electron values.

Notably, the nitrides spontaneous polarization obtained by Bernardini et al.[14] using Berry phase approach were: AlN: -0.081 C/m$^2$ (-5.05 x 10$^{-3}$ e/Å$^2$), GaN: -0.029 C/m$^2$ (-1.81 x 10$^{-3}$ e/Å$^2$), and InN: -0.032 C/m$^2$ (-1.99 x 10$^{-3}$ e/Å$^2$). These results are slightly lower that our results obtained from dipole calculations. Additional difference stems from the fact that, the dipole formulation presented above, gives two different polarization values.

## V. POLARIZATION INDUCED FIELD IN BULK AlN, GaN AND InN.

In addition to the polarization, the physically relevant quantity, directly affecting the performance of MQW based devices, is polarization induced electric field. Finite systems could be subjected to arbitrarily selected boundary conditions for potential, giving rise to different field inside. A standard case of a flat parallel plate capacitor demonstrates the dilemma. For the plates uniformly charged, the solution of Poisson equation having zero field outside is routinely selected by invoking additional argument that the potential should be finite at infinity. Formally, the second solution, that of the zero field inside and uniform nonzero field outside, fulfills Poisson equation as well as any normalized linear combination of these two solutions. Thus a plethora of the solutions exists, in which an additional argument of potential at infinity is applied to find physically sound solution. . In fact, this freedom is used in solution of Poisson equation by FFT method. Nevertheless, the physically sound solution is found using the same criterion as for capacitor with zero field at infinity. This solution was obtained for zero density outside, i.e. for the two selected cases above. This formulation allows obtaining the two possible values of the field, which corresponds to two selection of the crystal termination. In Fig. 5, the electric fields obtained for selected cases shown in Fig.2. The physically sound solutions are plotted using solid lines.

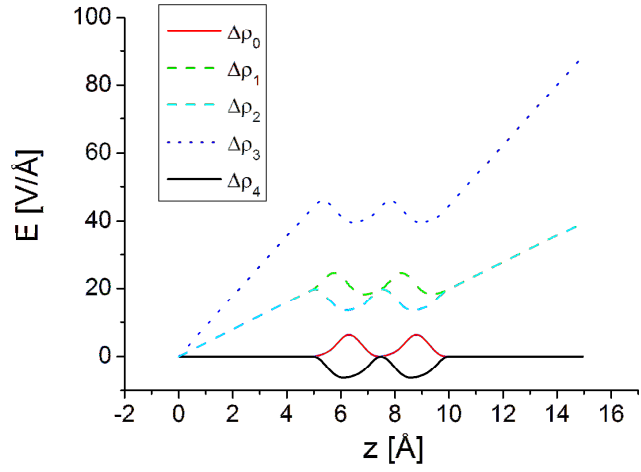

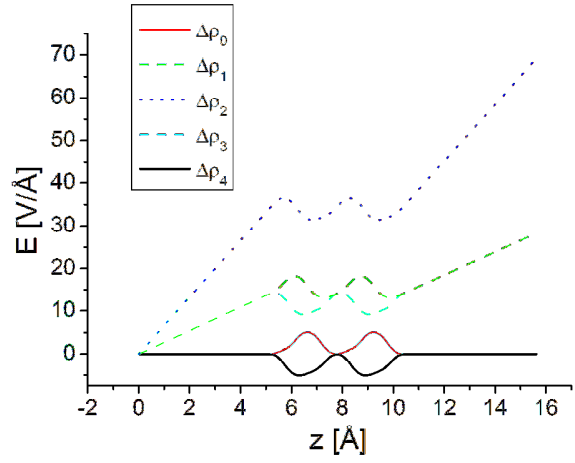

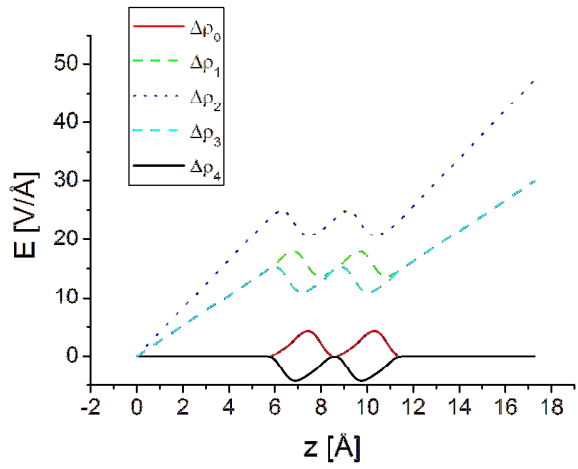

Fig. 10. (Color online) Electric field, averaged in c-plane, along c-axis, in single periodic cell of: (a) AlN; (b) GaN; (c) InN.

In addition, Fig. 11 presents the physically sound electric potential distributions plotted for AlN, GaN and InN. From these potential distributions the following potential differences were obtained: for AlN: $\Delta V^N_{AlN}$ = -18.00 V and $\Delta V^{Al}_{AlN}$ = 13.59 V; for GaN: $\Delta V^N_{GaN}$ = -14.49 V and $\Delta V^{Ga}_{GaN}$ = 12.06 V; for InN: $\Delta V^N_{InN}$ = -12.85V and $\Delta V^{In}_{InN}$ = 11.75 V. Since the following lattice constants were adopted for the modeling: $c_{AlN}$ = 4.983 Å, $c_{GaN}$ = 5.206 Å and $c_{InN}$ = 5.756 Å, the average fields are as follows: for AlN: $E^N_{0,AlN}$ = -3.612 V/Å and $E^{Al}_{0,AlN}$ = 2.727 V/Å; for GaN: $E^N_{0,GaN}$ = -2.783 V/Å and $E^{Ga}_{0,GaN}$ = 2.317 V/Å; for InN: $E^N_{0,InN}$ = -2.232 V/Å and $E^{In}_{0,InN}$ = 2.041 V/Å. These values are relatively high and they should be relatively easy to detect in nitride based structures. Note that screening could decrease or remove its influence, in the case when the size of the structures is comparable or larger than the characteristic screening lengths.

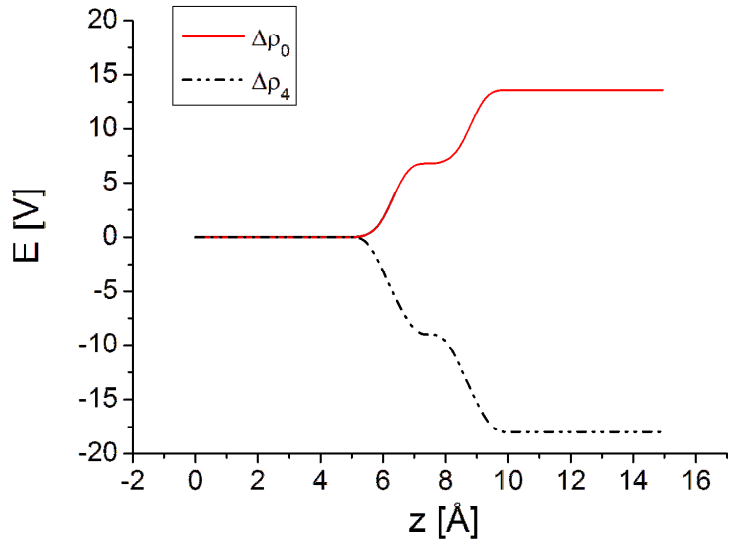

(a)

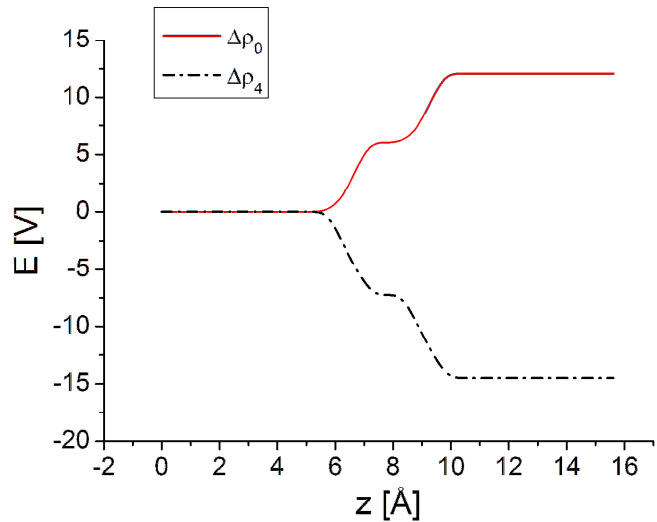

(b)

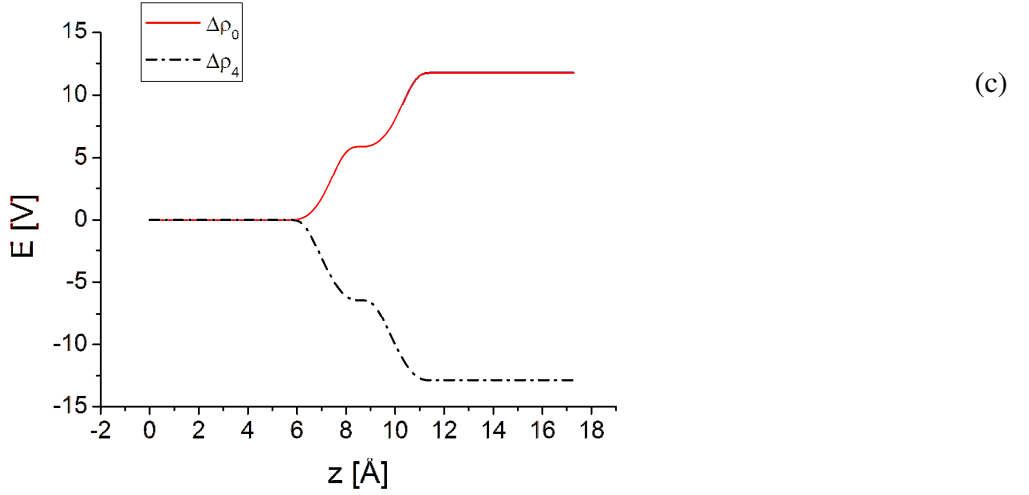

Fig. 11. (Color online) Electric potential, averaged in c-plane, along c-axis, in single periodic cell of: (a) AlN; (b) GaN; (c) InN. Black (solid) and red (broken) lines correspond to two different possible electric field directions, related to different polarization values.

These fields are to be used from in any geometrical arrangement of nitrides, encountered in electronic or optoelectronic devices by minimization of the electrostatic energy functional:

$$\Delta W = \int d^3 r \frac{\varepsilon (\vec{E} - \vec{E}_0)^2}{2} \quad (5)$$

Naturally, the uniform polarization field $\vec{E}_o$ is the minimal energy solution, corresponding to uniform polarization in the single finite size crystal without any surface or external contributions. These fields are the maximal field induced in finite size polarized semiconductors. In any real quantum structure, the fields should be lower, nevertheless that should affect properties of these structures considerably.

## VI. SUMMARY

Two different approaches to polarization of nitride semiconductors were assessed. It was shown that Berry phase formulation of the electron related polarization component provides a nonzero polarization for nonpolarized system. The electronic part gives saw-like pattern for polarization. Additionally a number of various solutions, different for various selection of the simulated volume could be obtained. A total number of these solutions, related to well known scaling of the geometric phase is equal to the number of valence electrons in the system. Summed with similar pattern for ionic part, provides several polarization values.

Standard dipole density formulation depends on the selection of the simulation volume in periodic continuous way. Using the condition of continuous embedding into the infinite medium, and simultaneously, the zero surface charge representation at crystal boundary, physically sound solution could be identified. This solution corresponds to maximal and minimal polarization values and corresponds to different physical termination of the crystal surfaces, either bare or covered by complementary atoms. This change leads to polarization and electric field reversal. Values of the fields are maximal values possible for finite size polarized nitrides without any surface charges or externals fields.

## ACKNOWLEDGEMENTS

The calculations reported in this paper were performed using computing facilities of the Interdisciplinary Centre for Modelling of Warsaw University (ICM UW). The research was partially supported by the European Union within European Regional Development Fund, through grant Innovative Economy (POIG.01.01.02-00-008/08). We would like to thank Michael Springborg from University of Saarland and Bernard Kirtman from University of California Santa Barbara for p discussion of importance of boundary conditions for polarization determination which has changed the present publication considerably.